# Evaluation methodology in the REVIT project

# Contents

**Authors**:

Anna Mavroudi , anna.mavroudi@st.ouc.ac.cy , Open University of Cyprus

Thanasis Hadzilacos, thh@ouc.ac.cy , Open University of Cyprus

**Language**: English

**Abstract**: The purpose of this technical report is to present a methodological approach for evaluating the REVIT educational venture as a whole and assessing the learning effectiveness of the resulting collaborative e-courses.

## Evaluation and assessment: 360 degrees feedback

Evaluation in an educational context has been defined by numerous scholars. The definition adopted here is that evaluation is "the collection of, analysis and interpretation of information about any aspect of a program of education or training as part of a recognized process of judging its effectiveness, its efficiency and any other outcomes it may have"(Ellington, Percival and Race, 1988). This applies both to the formative evaluation (i.e. collect data and information in order to improve the educational program, while the program is still developed) and to summative evaluation, which is conducted after formative evaluation completed and the implementation of the educational program is accomplished (Chen, 2011). On the other hand, assessment refers to "measuring learner performance.

Thus, it might be part of the evaluation, but evaluation and assessment are not synonymous (Lockee et al., 2002). Typically, formative evaluation in instructional design for distance education involves the following stages (Lockee et al., 2002; Smith & Ragan, 1999):

- design review (i.e. the use of other designers to evaluate the elements of design)
- expert review (i.e. the use of learning content experts, instructors, test creators and others to judge the appropriateness of the learning content and learning activities for the target learners)
- one-to-one review (i.e. the co-working of a member of the design/development team and a member of the learner population on a piece of the interface mockup of the distance education course)
- small group reviews (i.e. the co-working of a small group of the learner population on a fairly finished part of a distance education course)
- field trials (i.e. trying out the completed distance education course with a sample or samples of students in the actual field setting)
- ongoing reviews (i.e. the collection of data and identification of problems while the course is delivered)

From the types of formative evaluation mentioned above, the Revit evaluation process included: design review, expert review and ongoing reviews. Expect from the former (since the focus of this paper is away from the design) the other two types are described in the next sections.

### What exactly was evaluated in the REVIT project

*The Revit project as a whole*: This summative evaluation tries to respond to the main issue of the appropriateness of the project methodology. The "typical" aspects of the whole project, i.e. the schedules, deadlines and feasibility of the whole project were evaluated in order to notice eventually delays and retardations or other project specific problems.

*The communication (digital) tools:* in order to facilitate the communication between the partners, several digital tools were used, such as the webconference platform and various tools included in the Revit Distance Learning Services System (blogs, fora etc). Their (communicative) efficiency, overall utility, ease of use (consistency, compatibility, adaptability of the user interface), reliability and documentation was evaluated.

*Didactical value of the e-courses:* This was an important and complex part of the evaluation. Several web 2.0 tools were used in the e-courses in synchronous and asynchronous mode, for young people and adults, for several different themes. So, it is important to fully understand the didactical value of the e-courses. The improvement of the quality of the learning taking place, their ease of use and the added (learning) value were assessed, in order to establish an accurate description of the efficiency of these courses.

It should be noted here that there have been reported in the recent bibliography some cases of providing e-courses to citizens of rural areas for professional development purposes, like teacher training (Koulouris & Sotiriou, 2009). What actually differentiates REVIT is the integration of cost-effective (yet well-known for their educational affordances) ICTs with the combined use of the already available infrastructure in the rural areas in order to provide educational opportunities through distance learning.

*Face-to-face project meetings*: Some face-to-face meetings were planned (kickoff, intermediates, final). Each one of them was also evaluated, since participants were invited to answer in prepared questionnaires and their responses were analyzed.

## The evaluation instruments

To obtain feedback from the e-tutors, a reflective questionnaire (annex 4) was emailed, to be used complementarily with the other means of evaluation:

- the tutor diary (completed by tutors after each synchronous web-conference session and used as a means of formative evaluation; see annex 1)
- the evaluation report (completed by tutors at the end of the e-course and used for summative evaluation; see annex 2) and
- the trainees' questionnaire (completed by the learners at the end of the e-course and was used as a means of summative assessment, see annex 3).

The tutor's diary aimed mainly at conducting evaluation at the lesson level, whereas the reflective questionnaire along with the evaluation report aimed at the course level. The online trainees' questionnaire (in the form of anonymous online survey) aimed to reveal strengths and weaknesses not only specific to the teaching-learning process, but also course-specific parameters, like: relevancy of the learning material, user-friendliness of the e-learning platform graphical interface etc.

Finally, focus groups were also realised, which, in the Revit project, could be categorized as:

- country focus groups (participants from each country) and
- role focus groups (i.e one focus group for learners and one focus group for teachers).

In a focus group interview the idea was to find a common understanding related to the issue at stake, with the guidance and support of a moderator. The focus groups interviews in the Revit project were organized either on a face-to-face base or through the digital platform of synchronous communication. Some guidelines were given by the external evaluator beforehand the sessions. The role of moderator was usually held by a person who was already familiar with the participants and the project. This was helpful in making the participants feel comfortable in expressing their opinions freely. This person was a member of the research team who originated from an application area, a tutor of an e-course, the external evaluator etc.

## Annex 1: The tutor's diary notes

This diary was completed by the e-tutor after the end of each of the synchronous session and was used as a means of formative evaluation.  For each lesson, the following data were gathered:

1. Date of the session  and duration

2. Number of participants

3. Location of teacher and students (e.g. teacher in Patras, Greece  3 students in Palaichori, Cyprus and 4 Students in Ios, Greece)

4. Content (e.g. "Introduction to the use of blogs")

5. Contribution-activities of the learners (during the session or homework e.g. *the learners participated in a discussion about ….* **Or** *learners had to prepare a text and then to upload it* **or** *the learners had to create a facebook account for the next session…* etc)

6. Technical problems (e.g. *problems of sound, system collapsed, session not recorded or partially recorded* …)

7. Any facts of special interest (e.g. "*students weren't very active because* …."or "*students had a discussion about the value of the course*…" or "*students had a discussion about their personal  interests"*  or any event  that could be useful in order to better understand their "private relation" - their real, personal relation - to the course in general, such as feelings etc)

## Annex 2: The evaluation report

The evaluation report was also addressed to the e- tutor, it was completed at the end of the course (i.e.  a series of lessons) and it was used as part of summative evaluation. For each course it was important to know:

1. Title and dates of the course. A general, short description of its content and its goals.

2. A general, short description of the learning scenarios, i.e. a description of the organization of a "typical" e-course (e.g. an introduction to the main subject for 5', then a discussion with participants, activities for 20', etc)

3. Number of participants and location of participants and moderator. A (short) description of the profile of the participants (age, profession or socio-economic status, education, ICT background, including previous experience with seminars or online courses)

4. A description of the planned activities of students (e.g. what kind of homework? creation of a personal blog, creation of a facebook profile during the session,  etc).

5. Remarks on drop outs (if any) along with a justification.

6. Tutor's about the platform used and a description of persistent technical problems (if any)

7. Tutor's opinion about the management of the class. Whether it was satisfactory – compared to his/her expectations and compared to the management of a face-to-face classroom. Also, the tutor is asked to reflect whether time management was satisfactory (frequency of occurrence of very short or very long sessions etc).

8. The tutor is asked to express his/her opinion about the achievement of the didactic goals of the course. (Did he/she face any particular problems? Were the learners well prepared for this course? etc) and whether he/she thinks that distance teaching/learning is an appropriate "vehicle" for this kind of course.

9. Remarks or comments on the REVIT methodology (e.g. organization of the courses, method of courses development etc).

10. The tutor argues whether they were any "collateral" positive (or negative) didactic effects, e.g. more self-confidence, familiarity with computer and ICT in general etc.

### Annex 3: OnlineTrainees' questionnaire

#### Introductory questions
Question 1. Your responses concern the following program (Introductory Course on using Web Technologies, Tourism focused English Language course, E-Marketing Local Products and services/E-Commerce, Rural Tourism, Organic Farming, EFL Teachers using Web Technologies, Social media for teachers, Special education needs course, Environmental awareness for pupils, Successful family functioning)

Question 2. Country from which you attended the course (BG, CY, GR, FIN, PL)

#### Demographic data
Question 3. Sex (Female, Male)

Question 4. Age (No more than 30, Over 30 up to 40, Over 40 up to 50, Over 50 up to 60, Over 60 years old)

Question 5. Profession (Teacher, Agriculture, Small Business, Housework, other)

Question 6. Education (Primary, Secondary, Tertiary, Other)

Question 7. Is this course your first distant course? (Yes/No)

#### Quality of REVIT courses
Question 8. You were sufficiently prepared (in terms of the needed prerequisites) to follow this REVIT course (answers in a 5-point Likert scale)

Question 9. The distance courses of REVIT were satisfactory and helped you to achieve your personal goals (answers in a 5-point Likert scale)

Question 10. The distance courses of REVIT were rather hard to follow (answers in a 5-point Likert scale)

Question 11. The environment (interface) where the REVIT e-courses have been presented was rather hard to get familiar with and to use it (answers in a 5-point Likert scale)

Question 12. What was the frequency of the synchronous meetings of the REVIT e-course

(1 meeting every week, 2 meetings every week, 3 meetings every week, No answer)

Question 12a. Please choose the frequency that you prefer

(1 meeting every week, 2 meetings every week, 3 meetings every week, 4 meeting every week, No answer)

Question 13. The comments on your assignments and homework was satisfactory and helped you (answers in a 5-point Likert scale)

Question 14. The assignments and homework proposed by the teachers were useful and not hard to accomplish (answers in a 5-point Likert scale)

Question 15. Your tutors were well prepared for the lessons and eager in providing assistance towards the accomplishment of your learning goals (answers in a 5-point Likert scale)

### The didactic material and the platform used

Question 16. The electronic delivery of the course material (in digital form) was satisfactory (answers in a 5-point Likert scale)

Question 17. The (didactic) material used in the REVIT courses was well structured and useful (answers in a 5-point Likert scale)

Question 18. In general, it is easier to work with printed didactic material, than digital one and you prefer the printed material (answers in a 5-point Likert scale)

### The value of Web 2.0 Tools

Question 19. Blogs were an important element for your training (answers in a 5-point Likert scale)

Question 20. Wikis were an important element for your training (answers in a 5-point Likert scale)

Question 21. Podcasts were an important element for your training (answers in a 5-point Likert scale)

Question 22. The interaction with the platform used for your e-courses (moodle) was satisfactory (answers in a 5-point Likert scale)

Question 23. What extra feature would you suggest? (open-ended question)

### Technical aspects

Question 24. The following technical problems were important and had an influence on the courses: Technical problems related to my computer, Technical problems of the network (e.g.

low speed connection), Technical problems of the platform of communication (i.e. system e-lluminate), No technical problems

Question 25. It was necessary to have some equipment (i.e. computers and Internet access) in order to accomplish the assignments and homework proposed by the teachers (answers in a 5-point Likert scale)

### Social and Communicative aspects
Question 26. During the e-courses you met new people (your classmates) and you have developed some friendship with them (answers in a 5-point Likert scale)

Question 27. During the communication with your teacher it was rather easy (and you felt comfortable) to ask questions (answers in a 5-point Likert scale)

## Annex 4: The reflective questionnaire
In order to assess the learning effectiveness of the educational techniques used in the e-courses, the tutors reflected on the following techniques that they have used during the synchronous e-learning sessions (i.e. webconferences).  Focus was placed particularly in the assessment of two techniques that were used by almost all tutors: discussions and mini-lectures. More specifically, the reflective questionnaire was comprised of the following prompts:

-    The technique of mini-lecture used in the web-conference
The tutor was asked to mention three strong and three weak points concerning the technique of mini lecturing (with concurrent slide presentation) that was used in the web-conferences during the REVIT e-course. The tutor expressed his/her opinion on what can be done about the weak points.

-    The technique of discussion used in the web-conference
 The tutor was asked to think of the discussions throughout the web conferences and reflect whether they were satisfactory. Thought-provoking questions might be helpful. For example, questions like "How often do you estimate that a discussion took place: (a) among the learners, and (b) between learners and tutor? Did everybody speak? Did all/most students exchange experiences and ideas through dialogue? Were there any technical problems hindering the dialogue? How easy/difficult was it to coordinate the discussion? Compare with corresponding circumstances in face-to-face (traditional) lessons. In which part of the lesson did discussions usually take place, (a) introduction:  preparation -connection with previous (prerequisite) knowledge,  (b) main part of the lesson: presentation of the new knowledge,  (c) closing: review and assignment of (asynchronous) homework, (d) other "

-    Other techniques used
The tutor was asked to select one or two of the techniques that he/she has used in the learning process (such as: diagnostic questionnaire, collaborative authoring through the wiki, asynchronous dialogue using the forum etc). In order to evaluate their effectiveness in their context of use (that is, the REVIT e-courses) the tutor was prompt to consider issues such as the following:

- the degree in which the educational goals of the e-course were actually accomplished. In order to answer, the tutor was advised to reflect on the learners participation (i.e. active? frequent? onerous?  and if so, what kind of difficulties?) and on their performance.

-  whether another technique be more appropriate or better and if so, to explain why (Would it be easier? More interesting/attractive? Etc)

-  whether the tutor could  think of other educational contexts in which these techniques would be suitable If any) and to provide an example/scenario.

- Whether the tutor has thought of other ways that would have been helpful in the creation of a collaborative e-class

After having answered the above, the tutor then examines the results of the evaluation reports of the REVIT e-courses already completed. Having read the evaluation reports, he/she tries to recognise weaknesses common between his/her course and other courses. Finally, the tutor makes specific proposals on how these weaknesses could be overcomed in future e-courses.